\documentclass[twocolumn,english,prl,preprintnumbers,amsmath,amssymb]{revtex4}
\usepackage[T1]{fontenc}
\usepackage[latin9]{inputenc}
\usepackage{amstext}

\usepackage{amsmath}
\usepackage{amsfonts}
\usepackage{amssymb}
\usepackage{mathtools}
\usepackage{graphicx}
\usepackage {bm}

\makeatletter
\@ifundefined{textcolor}{}
{%
 \definecolor{BLACK}{gray}{0}
 \definecolor{WHITE}{gray}{1}
 \definecolor{RED}{rgb}{1,0,0}
 \definecolor{GREEN}{rgb}{0,1,0}
 \definecolor{BLUE}{rgb}{0,0,1}
 \definecolor{CYAN}{cmyk}{1,0,0,0}
 \definecolor{MAGENTA}{cmyk}{0,1,0,0}
 \definecolor{YELLOW}{cmyk}{0,0,1,0}
}

\def\be{\begin{equation}}
\def\ee{\end{equation}}
\def\bea{\begin{eqnarray}}
\def\eea{\end{eqnarray}}
\def\bse{\begin{subequations}}
\def\ese{\end{subequations}}

\usepackage{dcolumn}% Align table columns on decimal point
\usepackage{bm}% bold math
\makeatother

\usepackage{babel}
\begin{document}

\preprint{\bibliographystyle{revtex4}}

\title{Contrasting work fluctuations and distributions in systems with short-range and long-range correlations}

\author{T.R. Kirkpatrick$^{1}$, J.K. Bhattacherjee$^{1,2}$, and J.V. Sengers$^{1}$}

\affiliation{$^{1}$Institute for Physical Science and Technology, University of Maryland, College Park, MD 20742, USA\\
 $^{2}$Harish-Chandra Research Institute, Allahabad 211019, India}

\date{\today}
\begin{abstract}
It is shown that the work fluctuations and work distribution functions are fundamentally different in systems with short-range versus long-range correlations. The two cases considered with long-range correlations are magnetic work fluctuations in an equilibrium isotropic ferromagnet, and work fluctuations in a non-equilibrium fluid with a temperature gradient. The long-range correlations in the former case are due to equilibrium Goldstone modes, while in the latter they are due to generic non-equilibrium effects. The magnetic case is of particular interest since an external magnetic field can be used to tune the system from one with long-range correlations, to one with only short-range correlations. It is shown that in systems with long-range correlations the work distribution is extraordinarily broad compared to systems with only short-range correlations.
Surprisingly these results imply that fluctuations theorems such as the Jarzynski fluctuation theorem are more useful in systems with long-range correlations than in systems with short-range correlations.

\end{abstract}

\maketitle
In recent years there has been an enormous amount of research on some of the fundamental aspects of thermodynamics, engine efficiencies, and especially on so-called fluctuation theorems \cite{Buchkov_Kuzovlev_1977, Evans_Searles_1994, Gallavotti_Cohen_1995, Jarzynski_1997, Esposito_Harbola_Mukamel_2009, Campisi_Hanggi_Talkner_2011, Shiraishi_et_al_2016}. One of the central quantities considered is the thermodynamic work, its fluctuations, and the complete work distribution. For example, the Jarzynski fluctuation theorem (JFT) \cite{Jarzynski_1997} is $\langle e^{-\beta W}\rangle=e^{-\beta\Delta F}$ where $\beta=1/(k_{B}T)$, $W$ is the work, the angular brackets denote an average over a work distribution from one thermodynamic state to another, and $\Delta F$ is the free energy difference between the two states. In \cite{Kirkpatrick_Dorfman_Sengers_2016} it was shown that the fluctuations in the JFT are so large for systems with short-range correlations,  that they invalidate its practical use for all but the smallest systems.

In the same paper \cite{Kirkpatrick_Dorfman_Sengers_2016} it was also shown that the work distribution in a non-equilibrium steady state (NESS) with long-range correlations was very anomalous compared to a system with only short-range correlations. In particular, it was shown that the work distribution is very broad for systems with long-range correlations. Among other things, this implies that the JFT is more useful in systems with long-range correlations than in systems with only short-range correlations. Technically this is because a broad work distribution provides more support for quantities, such as $e^{-\beta W}$, determined by the tails of the distribution than a sharply peaked distribution with very small tails.
Here we explore this important point more generally by contrasting the work distribution function in systems with long-range correlations to the work distribution in systems with only short-range correlations. For this purpose we will compare and contrast the work distribution in an equilibrium isotopic ferromagnet, where there are long-range correlations due to Goldstone's theorem \cite{Chaikin_Lubensky_1995, Forster_1975}, and a non-equilibrium fluid in a temperature gradient where there are generic long-range correlations \cite{Kirkpatrick_Cohen_Dorfman_1982A, Dorfman_Kirkpatrick_Sengers_1994, DeZarate_Sengers_2006}, to the work distribution in systems with only short-range correlations. We generally find that in the long-range case the distribution is very broad compared to the short-range case and that, for fixed system size, its weight near the origin is suppressed compared to the short-range case. 

For the magnetic case we assume a single three-dimensional ferromagnetic domain that is ordered in the $z$-direction. To be specific we assume the domain to exist in the region between $z=0$ and $z=L$, and that there is perfect ordering  in the $z$-direction at these boundaries. That is, the transverse magnetic fluctuations vanish at $z=0$ and $z=L$. We further assume periodic boundary conditions in the transverse direction with $L_{x}=L_{y}=L_{\perp}$, and that $L_{\perp}/L\gg 1$. If $h$ is the magnitude of an external magnetic field in the $z$-direction, and if we assume that an applied field does not change the system volume, then the differential fluctuating magnetic work can be defined by \cite{Callen_1985, Kittel_1958} \footnote{The differential magnetic work is also often defined as $d\widetilde{W}_{\mathrm{mag}}(\mathbf x)=hd\widetilde{m}_{z}(\mathbf x)$. These two choices are related by a Legendre transformation.} $d\widetilde{W}_{\mathrm{mag}}(\mathbf x)=-\widetilde{m}_{z}(\mathbf x)dh$, with $\widetilde{m}_z(\mathbf x)$ the fluctuating magnetization in the $z$-direction. For small magnetic field, the total fluctuating magnetic work is simply $\widetilde{W}_{\mathrm{mag}}=-L_{\perp}^2Lh\widetilde{m}_z(L)$, where $\widetilde{m}_z(L)$ denotes the spatial average of $\widetilde{m}_z(\mathbf x)$.

Here we are interested in the fluctuating magnetic work, $\widetilde{W}_{\mathrm{mag}}$, deep in the ferromagnetic phase where $m_z$ is given by the transverse magnetization fluctuations, $\bm {\pi}(\mathbf{x})$, as $\widetilde{m}_{z}(\mathbf{x})=m_{0}\sqrt{1-\bm{\pi}^2(\mathbf{x})/m_0^2}\approx m_0-\bm{\pi}^2(\mathbf{x})/2m_0$ \footnote{This quadratic approximation can be justified by retaining higher-order terms and verifying they do not qualitatively change any of our results.}. For small fields then $\widetilde{W}_{\mathrm{mag}}=L_{\perp}^2Lh{\bm{\pi}^2(L)}/2m_0$, where ${\bm{\pi}^2(L)}$ is the spatial average of ${\bm{\pi}^2(\mathbf{x})}$. The $\bm{\pi}$ fluctuations are of long-range at zero magnetic field due to Goldstone's theorem. In wavenumber space, where $\bm\pi(\mathbf{x})=\frac{2}{L}\sum_{N=1}\int_{\mathbf{k}_{\perp}}e^{i\mathbf{k_{\perp}}\cdot\mathbf{x}_{\perp}}\sin({\frac{N\pi z}{L}})\bm\pi(\mathbf{k})$, and at finite $h$ they are given by \cite{Chaikin_Lubensky_1995, Forster_1975},

\begin{equation}
\langle \pi_i(\mathbf k)\pi_j(\mathbf {-k})\rangle=\frac{LL_{\perp}^2\delta_{ij}k_BT}{Jk^2+h/m_0}.
\end{equation}
Here $i,j=(x,y)$, $J$ is related to the magnetic exchange interaction, $k_B$ is Boltzmann's constant, $T$ is the temperature, $k^2=k_{\perp}^2+k_z^2$, with $k_{\perp}=\sqrt{k_x^2+k_y^2}$ the transverse wavenumber and $k_z^2=N^2\pi^2/L^2$. At $h=0$ the $1/k^2$ dependence in Eq.(1) indicates long-range or power law correlations in real space.

For the non-equilibrium fluid case, we consider a fluid with a temperature gradient in the $z$-direction. 
The dimension of the system in the $z$-direction is $L$ while in the perpendicular direction it is $L_{x}=L_{y}=L_{\perp}$ and we again assume $L_{\perp}\gg L$. For most liquid systems the thermal conductivity varies little with temperature so we can assume a linear temperature profile given by,
\begin{equation}
T(z)=T_{0}+{\frac{\Delta T}{L}}z.
\label{eq:2.1}
\end{equation}
Here $\Delta T$ is the temperature difference between the two walls in the $z$-direction. In this case there are long-range temperature fluctuations, $\delta T(\mathbf{x})$. We again assume periodic boundary conditions in the transverse direction and perfectly conducting walls at $z=0$ and $z=L$ so that as a function of position $\delta T(\mathbf{x})$ exactly vanishes at the walls.

The long-range part of the local temperature fluctuations, $\delta T(\mathbf{x})$ are, in wavenumber space \cite{Kirkpatrick_Cohen_Dorfman_1982A, Kirkpatrick_Cohen_1983, Dorfman_Kirkpatrick_Sengers_1994, DeZarate_Sengers_2004, DeZarate_Sengers_2006}, $\delta T(\mathbf{x})=\frac{2}{L}\sum_{N=1}\int_{\mathbf{k}_{\perp}}e^{i\mathbf{k_{\perp}}\cdot\mathbf{x}_{\perp}}\sin({\frac{N\pi z}{L}})\delta T(\mathbf{k})$,
\begin{eqnarray}
\langle\delta T({\bf k})\delta T(-{\bf k})\rangle _{\mathrm{NESS}}=  \frac{LL_{\perp}^2k_{B}T}{\rho D_{T}(\nu+D_{T})}\frac{(k_{\perp}\nabla T)^{2}}{k^{6}}.\
\end{eqnarray}
Here $\rho$, $\nu$ and $D_{T}$ are the mass density, the kinematic
viscosity and thermal diffusivity of the fluid.  All of the thermo-physical parameters in Eq.(2.3) may be identified with their spatially averaged values \cite{Segre_et_al_1992}. We note that this correlation function is long ranged
as indicated by its $k^{-4}$ behavior at small wave numbers, while
the equilibrium temperature fluctuations are of very short range in space
with no singular behavior of the corresponding Fourier transforms
at small wave numbers. Note also, that since $\nabla T\propto 1/L$ the length scaling behavior of Eqs.(1) and (3) are identical for $h\rightarrow 0$. 

The important fluctuating contribution to the pressure in a NESS has been identified elsewhere \cite{Kirkpatrick_Dorfman_2015, Kirkpatrick_DeZarate_Sengers_2013} as $\widetilde{p}_{\mathrm{NE}}(\mathbf{x})=A[\delta T(\mathbf{x})]^{2}$ with $A=\rho(\gamma-1)[c_{p}-(\partial (c_{p}/\alpha)/\partial T)_{p}]/2T$. Here $c_{p},\gamma,\alpha$ are, respectively, the specific heat capacity at
constant pressure, the ratio of specific heat capacities, and the coefficient
of thermal expansion.
The fluctuating work in this case is given by $d\widetilde{W}_{\mathrm{NE}}=-\widetilde{p}_{\mathrm{NE}}(L)dV$ where $\widetilde{p}_{\mathrm{NE}}(L)$ is the spatial average of $\widetilde{p}_{\mathrm{NE}}(\mathbf x)$ \cite{Kirkpatrick_DeZarate_Sengers_2016a, Kirkpatrick_DeZarate_Sengers_2015, Kirkpatrick_Dorfman_Sengers_2016}. If the system expands in the $z$-direction from length $L$ to length $L(1+\Delta)$, and if $|\Delta|\ll 1$, then the fluctuating non-equilibrium work is simply $\widetilde{W}_{\mathrm{NE}}=-L_{\perp}^2L\Delta\widetilde{p}_{\mathrm{NE}}(L)$. To simplify our notation, so that both the magnetic and non-equilibrium work are positive, we actually consider a contraction and use $\Delta=-|\Delta|$.

We will first give the results and partially outline their derivation, and then discuss their applicability. Additional technical details will then be given.
From Eq.(1), it is obvious that the long-range aspect of the magnetic work distribution will only occur for small $h$. We find that the work cumulant for both cases, retaining only universal long-range contributions \footnote{The average magnetic  work is $\propto L_{\perp}^2L=V$ while the average NE work is $\propto L_{\perp}^2$. This is the only structural difference between the two examples, because the fluctuation properties of the two are identical. Here we ignore this difference, assume $\langle{W}\rangle_{\mathrm{mag}}=a_{\mathrm{mag}}b_{\mathrm{mag}}L_{\perp}^2$, and discuss later in Remark 2 the minor modifications needed in the results if the average work is $\propto V$.}, can then be written, setting $k_BT=1$,
\begin{equation}
\langle\widetilde{W}_{\alpha}^n\rangle_{\mathrm{cumulant}}\equiv\kappa_{\alpha}(n)= a_{\alpha}\frac{L_{\perp}^2}{L^2}(b_{\alpha}L^2)^ng_\alpha(n),
\end{equation}
with $\alpha=(\mathrm{mag}, \mathrm{NE})$, $a_{\mathrm{mag}}={\pi}/4$, $b_{\mathrm{mag}}=2h/(\pi^2m_0J)$, $g_{\mathrm{mag}}(n)=(n-2)!\zeta (2n-2)$ and $a_{\mathrm{NE}}={\pi}/8$, $b_{\mathrm{NE}}=8Al(\Delta T)^2|\Delta| /27{\pi}^4$, $g_{\mathrm{NE}}(n)=(27/4)^n\zeta (4n-2)n!(n-1)!(2n-2)!/(3n-1)!$. Here $l=(\rho D_{T}(\nu+D_{T}))^{-1}$ is a microscopic length and $\zeta(n)$ is the Riemann zeta function of order $n$. For $n\gg1$ we note $g_{\mathrm{mag}}/n!\approx 1/n^2$ and $g_{\mathrm{NE}}/n!\approx (\sqrt{3\pi}/2)/n^{3/2}$.

With Eq.(4) we can determine the work distribution, defined by $\rho_{\mathrm{\alpha}}(W)$,   $\alpha=(\mathrm{mag}, \mathrm{NE})$, as follows. First  we define a cumulant generating function, $K_{\alpha}(t)$ by,
\begin{equation}
K_{\alpha}(t)=\ln \Big(\int{dWe^{Wt}\rho_{\mathrm{\alpha}}(W)}\Big) =\sum_{n=1}\frac{\kappa_{\alpha}(n)t^n}{n!}.
\end{equation}
The work distribution is now formally given by the inverse transform,
\begin{equation}
\rho_{\mathrm{\alpha}}(W)=\int dte^{-Wt+K_{\alpha}(t)}.
\end{equation}
The integral in Eq.(6) can be evaluated using saddle-point or steepest-descent methods because the $\textit{scale}$ of $W$ grows with $L_{\perp}^2/L^2$. The important feature in the evaluation of Eq.(6) is the convergence property, or singularity structure, of $K_{\alpha}(t)$, Eq.(5), which in turn is determined by the large $n$-behavior of $\kappa_{\alpha}(n)$, given below Eq.(4).

Neglecting non-exponential pre-factors one generally finds,
\begin{equation}
\rho_{\alpha}(W)\propto e^{-a_{\alpha} \frac{L_{\perp}^{2}}{L^2}G_{\alpha}(\widehat{W}_{\alpha})}\theta(W),
\end{equation}
where $\widehat{W}_{\alpha}=W/a_{\alpha}b_{\alpha}L_{\perp}^2$ and $G_{\alpha}(\widehat{W}_{\alpha})=\widehat{W}_{\alpha}{K}_{\alpha}'^{-1}(\widehat{W}_{\alpha})-K_{\alpha}({K}_{\alpha}'^{-1}(\widehat{W}_{\alpha}))$ with ${K}_{\alpha}'^{-1}$ the inverse function of $K_{\alpha}'=dK_{\alpha}(t)/d(b_{\alpha}L^2t)$. The crucial aspect of Eq.(7) is that the function $G_{\alpha}$ does not explicitly depend on the system size, so that the scale of the exponential is $L_{\perp}^2/L^2$, while the scale of $W$ in the tails is $\propto L_{\perp}^2$. In detail, one finds for the  tails of the distributions,
\begin{equation}
\rho_{\alpha}({W}\rightarrow{0})\propto e^{-a_{\alpha, 0}\frac{L_{\perp}^2}{L^2}(\ln1/\widehat{W}_{\alpha})^{s_{\alpha}}}\theta({W}),
\end{equation}
and,
\begin{equation}
\rho_{\mathrm{\alpha}}({W}\rightarrow\infty)\propto e^{-a_{\alpha, \infty}\frac{L_{\perp}^2}{L^2}\widehat{W}_{\alpha}}\theta({W}).
\end{equation}
Here $s_{\mathrm{mag}}=2$, $s_{\mathrm{NE}}=3/2$, $a_{\mathrm{mag}, 0}=\pi/8$, $a_{\mathrm{NE}, 0}=\pi/4\sqrt{3}$, $a_{\mathrm{mag}, \infty}=\pi/4$ and $a_{\mathrm{NE}, \infty}=\pi/8$.  In between the tails the distribution functions can be taken to be Gaussian,
\begin{equation}
\rho_{\alpha}({W}\approx\langle W\rangle_{\alpha})\propto e^{-\frac{a_{\alpha, \mathrm{G}}L_{\perp}^2}{2L^2}(\overline{W}_{\alpha}-1)^2}\theta(W),
\end{equation}
where $\overline{W}_{\alpha}=W/\langle{W}\rangle_{\alpha}$, with $\langle{W}\rangle_{\alpha}$ the average work determined by $\rho_{\alpha}$, and $a_{\mathrm{mag}, \mathrm{G}}=3/2\pi$ and $a_{\mathrm{NE}, \mathrm{G}}=1575/64\pi$. These results are to be contrasted to those for a system with only short-range (SR) correlations. For example, for an ideal gas of $N$ particles undergoing a fractional volume change $\propto\epsilon=(1+\Delta)^{-2/3}-1> 0$, the equivalent results are \cite{Crooks_Jarzynski_2007},
\begin{equation}
\rho_{\mathrm{SR}}(W\rightarrow{0})\propto e^{-\frac{3}{2}N\ln(1/\overline{W})}\theta(W)
\end{equation}
and,
\begin{equation}
\rho_{\mathrm{SR}}(W\rightarrow\infty)\propto e^{-\frac{3}{2}N\overline{W}}\theta(W)
\end{equation}
and,
\begin{equation}
\rho_{\mathrm{SR}}(W\approx\langle W\rangle)\propto e^{-\frac{3}{2}N(\overline{W}-1)^{2}}\theta(W),
\end{equation}
with $\langle W\rangle=3N\epsilon/2$. Note that the pre-factor in the exponentials in Eqs.(11)-(13) scales as the system size, $N\propto L_{\perp}^2L$.

As an application of these results we have computed the fluctuations in the JFT. That is, with $\Omega=e^{-W}$ we consider the fluctuation measure,
\begin{equation}
\epsilon_{\Omega,\alpha}=\frac {\langle{\Omega}^2 \rangle_{\alpha} -{\langle\Omega \rangle_{\alpha}}^{2}}{{\langle\Omega \rangle_{\alpha}}^{2}}.
\end{equation}
With
\begin{equation}
\epsilon_{\Omega,\alpha}=e^{\frac{L_{\perp}^2}{L^2}F_{\alpha}(b_{\alpha}L^2)}
\end{equation}
we obtain, neglecting non-exponential pre-factors,
\begin{equation}
F_{\alpha}(b_{\alpha}L^2\ll 1)\approx c_{\alpha}(b_{\alpha} L^2)^2
\end{equation}
with $c_{\mathrm{mag}}=\pi^3/48$ and $c_{\mathrm{NE}}=9\pi^7/(2^9(175))$. We also obtain \footnote{The result given for this quantity in \cite{Kirkpatrick_Dorfman_Sengers_2016} was different because the work distribution given there was exact only for the region $\widehat{W}_{\mathrm{NE}}\gg 1$, and not the $\widehat{W}_{\mathrm{NE}}\rightarrow 0$ limit, which is the important part of the distribution for determining $F_{\mathrm{NE}}(b_{\mathrm{NE}}L^2\gg 1)$.} \footnote{If we use Eq.(4) for $b_{\mathrm{mag}}L^2\gg 1$, a calculation gives $F_{\mathrm{mag}}(b_{\mathrm{mag}}L^2\gg 1)\approx\pi(\ln(b_{\mathrm{mag}}L^2))^{2}/8$ in perfect accord with Eq.(17) for the NE fluid case. As noted in Remark 3, however, in this limit Eq.(4) must be modified for the magnetic case.},
\begin{equation}
F_{\mathrm{NE}}(b_{\mathrm{NE}}L^2\gg 1)\approx\frac{\pi}{4\sqrt{3}}(\ln(b_{\mathrm{NE}}L^2))^{3/2}.
\end{equation}
Note that because $b_{\alpha}L^2\ll 1$, Eq.(16) implies that the exponential factor in Eq.(15) is $\ll L_{\perp}^2/L^2$, while Eq.(17) implies that the exponential factor in Eq.(15) is, for $b_{\mathrm{NE}}L^2\gg 1$, logarithmically larger than $L_{\perp}^2/L^2$. For the SR case one obtains \cite{Kirkpatrick_Dorfman_Sengers_2016},
\begin{equation}
\epsilon_{\Omega, \mathrm{SR}}=e^{\frac{3}{2}N\ln(1+\frac{\epsilon^2}{1+2\epsilon})}.
\end{equation}
$\epsilon$ in Eq.(18) is analogous to $b_{\alpha}L^2$ in Eq.(15) and since $\epsilon_{\Omega, \mathrm{SR}}\approx e^{3N\epsilon^2/2}$ for $\epsilon\ll 1$ and $\epsilon_{\Omega, \mathrm{SR}}\approx e^{3N\ln\epsilon/2}$ for $\epsilon\gg 1$, we see these limiting cases are structurally like Eqs.(15)-(17). The obvious fundamental distinction between the long-range and short-range cases, is that in the former the scale of the exponential is $L_{\perp}^2/L^2$, while in the latter it is the system size or volume. We emphasize that while  the enormous fluctuations in the short-range case, for $N\gg 1$, restrict the utility of the JFT for such systems, our results imply that the JFT will be much more useful in systems with long-range correlations.

We next give some further technical details. We focus on the magnetic case, which is a bit simpler than the non-equilibrium fluid case. First, Eq.(4) follows from the Gaussian nature of the $\bm{\pi}$ fluctuations \cite{Belitz_Kirkpatrick_1997}, some simple combinatorics, and performing some elementary sums and integrals. Second, it is easy to show that the tails of the distribution are determined  by the large $n$ behavior of  $g_{\mathrm{mag}}(n)$ in Eq.(4). To that end we use the large $n$ result for $g_{\mathrm{mag}}(n)$ for all $n$. For the magnetic case, this allows us to sum the $t$ derivative of the cumulant generating function, $K_{\mathrm{mag}}(t)$. The saddle-point equation for $t$ in Eq.(6) is then,
\begin{equation}
W=\frac{dK_{\mathrm{mag}}(t)}{dt}=-\frac{a_{\mathrm{mag}}L_{\perp}^2}{L^2t}\ln(1-b_{\mathrm{mag}}L^2t).
\end{equation}
Note this equation only has a solution for $W\geq 0$. The solution for $W\rightarrow\infty$ is,
\begin{equation}
b_{\mathrm{mag}}L^2t\approx1-e^{-\widehat{W}_\mathrm{mag}}
\end{equation}
and the solution for $W\rightarrow 0$ is, here $t=-|t|$,
\begin{equation}
b_{\mathrm{mag}}L^2|t|\approx\frac{1}{\widehat{W}_\mathrm{mag}}\ln\frac{1}{\widehat{W}_\mathrm{mag}}.
\end {equation}
The Eqs.(8) and (9) for the magnetic work case. can then be obtained by integrating Eq.(19) for $b_{\mathrm{mag}}L^2|t|=-b_{\mathrm{mag}}L^2t\gg 1$ and $b_{\mathrm{mag}}L^2t\approx 1$, respectively. The Gaussian distribution, Eq.(10), follows from the small $t$ behavior of $K_{\mathrm{mag}}(t)$ and is fixed by the average magnetic work and it's fluctuations.

We conclude with a number of remarks:

\enumerate

\item Comparing Eqs.(7)-(10) with Eqs.(11)-(13) two things should be noted. First, because we focus on long-range fluctuating contributions to the work, the average work in Eqs.(7)-(10) scales as $\propto L_{\perp}^2$ and not like $N\propto L_{\perp}^2L$ as in Eqs.(11)-(13). This explains the numerators in the exponential factors in Eqs.(7)-(10). Second, the long-range nature of the correlations lead to the extra $1/L^2$ factors in these equations. That is, for the long-range case we have $\langle (W-\langle{W}\rangle_{\alpha})^2\rangle_{\alpha}\propto L^2\langle W\rangle_{\alpha} $, while for systems with short-range correlations the relationship is $\langle (W-\langle{W}\rangle_{\alpha})^2\rangle\propto\langle W\rangle$. The extra factor of $L^2$ indicates that the work distribution in system with long-range correlations is extraordinarily broad compared to the short range case.
\item As noted above in Remark 1, the average work in Eqs.(7)-(10) scales as $\propto L_{\perp}^2$, and not as the system volume. If the average work does scale as $V$, the fluctuations will still be determined by the long-range correlations. In this case the results summarized by Eqs.(7)-(10) are changed as follows. The pre-factor of the Gaussian, $L_{\perp}^2/L^2$, in Eq.(10) is replaced by $L_{\perp}^2$, and the $\overline{W}_{\alpha}$ in the Gaussian is $W$ normalized by the actual average work, $\propto V$. The tails of the distribution, however, are still controlled by the same pre-factors in the exponential $\propto L_{\perp}^2/L^2$, and are the same functions of $W$, normalized by $a_{\alpha}b_{\alpha}L_{\perp}^2$, as in Eqs.(8) and (9) \footnote{The argument of the $\theta$ function in Eqs.(7)-(10), and the $W$'s in Eq.(8), also gets replaced by $W-(\langle W\rangle_{\alpha}-c_{\alpha}a_{\alpha}b_{\alpha}L_{\perp}^2)$, with $c_{\alpha}$ a constant of order unity.}. The crossovers to the tail distribution occur when $|\overline{W}_{\alpha}-1|\approx O(1/L)$. Finally, the length and $b$ scalings in Eqs.(15)-(17) are unchanged.
\item The dimensionless parameter characterizing the magnetic field in Eq.(1) is $b_{\mathrm{mag}}L^2$ so that in taking the $b_{\mathrm{mag}}L^2\gg 1$ limit, a finite field must be taken into account there, as well as in integrating $d\widetilde{W}_{\mathrm{mag}}(\mathbf x)=-\widetilde{m}_{z}(\mathbf x)dh$. In the calculations this leads to a factor of  $(b_{\mathrm{mag}}L^2)^{3/2-n}$ in Eq.(4). The important result is that every term in Eq. (4) is $\propto L_{\perp}^2Lh^{3/2}=Vh^{3/2}$. The work distribution in this case is in the short-range universality class, with a non-analytic field dependence that reflects the long-range correlations at zero field. Also of interest is Eq.(14) for this case: $\epsilon_{\Omega, \mathrm{mag}}=e^{cL_{\perp}^2L(h/m_0J)^{3/2}}$, with $c$ a number of order unity. 

Physically all of this is obvious: For finite $h$ the correlations implied by Eq.(1) are of short-range, so that one expects the pre-factors in the work distribution to scale as the system volume, just as they do in Eqs.(11)-(13). The $h^{3/2}$ follows from the fact that the longitudinal magnetic correlations in a three-dimensional isotropic ferromagnet scale as \cite{Brezin_Wallace_1973, Belitz_Kirkpatrick_1997} $\chi_{L}(h\rightarrow{0})\propto 1/h^{1/2}$ and in the finite field magnetic work fluctuation calculation, this result is integrated twice.
\item Similar results are expected in other systems with long-range correlations, no matter the source of the correlations. Of particular interest are biological or electronic  and spintronic systems. For example, in active matter or living systems, various types of broken symmetries and Goldstone modes have been discussed in the literature \cite{Ramaswamy_2010, Marchetti_et_al_2013}.  Similarly, in electronic and spintronic systems, long-range correlations can arise from, for example, various types of magnetic order, or exist even more generically at low or zero temperatures \cite{Belitz_Kirkpatrick_Vojta_2005}.
\endenumerate

\medskip
Discussions with Bob Dorfman are gratefully acknowledged. This work was supported by the National Science Foundation under Grant
No. DMR-1401449. The research of JKB was supported by the Burgers Program for Fluid Dynamics of the University of Maryland and by an  APS-IUSSTF professorship award.

\end{document}